%% file: main.tex
\documentclass[a4paper,UKenglish]{lipics-v2016} 

\input{style}
\input{macros}
\usepackage{booktabs}
\usepackage{stmaryrd}
\usepackage{bussproofs}
\usepackage{mdwlist}
\usepackage{xspace}
\usepackage{xcolor}
\begin{document}

%% \special{papersize=8.5in,11in}
%% \setlength{\pdfpageheight}{\paperheight}
%% \setlength{\pdfpagewidth}{\paperwidth}

\title{Modelling homogeneous generative meta-programming}
\titlerunning{Modelling homogeneous generative meta-programming}

\author[1]{Martin Berger}
\author[2]{Laurence Tratt}
\author[2]{Christian Urban}

\affil[1]{University of Sussex}
\affil[2]{King's College London}
\authorrunning{M.\,Berger, L.\,Tratt, C.\,Urban}

\Copyright{Martin Berger, Laurence Tratt, Christian Urban}

\subjclass{D.3.3 Language Constructs and Features.}
\keywords{Formal Methods, Meta-Programming, Operational Semantics, Types, Quasi-Quotes, Abstract Syntax Trees.}

%Editor-only macros:: begin (do not touch as author)%%%%%%%%%%%%%%%%%%%%%%%%%%%%%%%%%%                                          
\EventEditors{}
\EventNoEds{2}
\EventLongTitle{}
\EventShortTitle{ECOOP 2017}
\EventAcronym{ECOOP}
\EventYear{2017}
\EventDate{December 24--27, 2016}
\EventLocation{}
\EventLogo{}
\SeriesVolume{42}
\ArticleNo{23}
% Editor-only macros::end %%%%%%%%%%%%%%%%%%%%%%%%%%%%%%%%%%%%%%%%%%%%%%%  

%% \authorinfo{}{}{} % for anonymous reviewing

\maketitle

\input{abstract}

\title

\newcounter{line}
\setcounter{line}{1}

\input{introduction}

\input{language}

\input{generalFramework}
\input{typing}

\input{conclusion}

\bibliography{bib}

\appendix

\end{document}

%% file: style.tex
\usepackage{microtype}%For ECOOP. if unwanted, comment out or use option "draft" 
\usepackage{amsmath}
\usepackage{amssymb}
\usepackage{booktabs}
\usepackage{url} 
\usepackage{ifthen}
\usepackage{listings}
\usepackage{graphicx} 
\usepackage{arydshln}

%% file: macros.tex
\usepackage{color}
\newboolean{showcomments}
\setboolean{showcomments}{true}
\ifthenelse{\boolean{showcomments}}
  {\newcommand{\mynote}[2]{
    \textcolor{red}{%
      \fbox{\bfseries\sffamily\scriptsize#1}
      {\small$\blacktriangleright$\textsf{\emph{#2}}$\blacktriangleleft$}
    }
   }
  }
  {\newcommand{\mynote}[2]{}
  }

\newcommand{\AXIOM}[2]{\AxiomC{$#1$}\RightLabel{\SMALLRULENAME{#2}}}
\newcommand{\UNARYINFERENCE}[2]{\UnaryInfC{$#1$}\RightLabel{\SMALLRULENAME{#2}}}
\newcommand{\BINARYINFERENCE}[2]{\BinaryInfC{$#1$}\RightLabel{\SMALLRULENAME{#2}}}
\newcommand{\TRINARYINFERENCE}[2]{\TrinaryInfC{$#1$}\RightLabel{\SMALLRULENAME{#2}}}

\newcommand{\TYPEDGENSYM}[1]{\mathsf{gensym}(\alpha)}

\newcommand{\LIFT}[1]{\mathsf{lift}(#1)}

\newcommand{\LETCT}[3]{\mathsf{let_\downarrow} #1 =\ #2\ \mathsf{in}\ #3}

\newcommand{\NI}{\noindent}

\newcommand{\VEC}[1]{\tilde{#1}}

\newcommand{\CODE}{\mathsf{Code}}
\newcommand{\TYPEDCODE}[1]{\mathsf{Code}\langle #1 \rangle}

\newcommand{\infer}[2]{\frac{\displaystyle{ #1 }}{\displaystyle{ #2 }}}
\newcommand{\ZEROPREMISERULE}[1]{\infer{}{#1}}
\newcommand{\ONEPREMISERULE}[2]{\infer{#1}{#2}}
\newcommand{\TWOPREMISERULE}[3]{\infer{#1 \quad #2}{#3}}
\newcommand{\THREEPREMISERULE}[4]{\infer{#1 \quad #2 \quad #3}{#4}}
\newcommand{\FOURPREMISERULE}[5]{\infer{#1 \quad #2 \quad #3 \quad #4}{#5}}
\newcommand{\FIVEPREMISERULE}[6]{\infer{#1 \quad #2 \quad #3 \quad #4 \quad #5}{#6}}

\newcommand{\RULENAME}[1]{\textsc{#1}}
\newcommand{\SMALLRULENAME}[1]{\textsc{\tiny #1}}
\newcommand{\ZEROPREMISERULENAMEDRIGHT}[2]{\ZEROPREMISERULE{#1}\,\SMALLRULENAME{#2}}
\newcommand{\ONEPREMISERULENAMEDRIGHT}[3]{\ONEPREMISERULE{#1}{#2}\,\SMALLRULENAME{#3}}
\newcommand{\TWOPREMISERULENAMEDRIGHT}[4]{\TWOPREMISERULE{#1}{#2}{#3}\,\SMALLRULENAME{#4}}
\newcommand{\THREEPREMISERULENAMEDRIGHT}[5]{\THREEPREMISERULE{#1}{#2}{#3}{#4}\,\SMALLRULENAME{#5}}
\newcommand{\FOURPREMISERULENAMEDRIGHT}[6]{\FOURPREMISERULE{#1}{#2}{#3}{#4}{#5}\,\SMALLRULENAME{#6}}
\newcommand{\FIVEPREMISERULENAMEDRIGHT}[7]{\FIVEPREMISERULE{#1}{#2}{#3}{#4}{#5}{#6}\,\SMALLRULENAME{#7}}

\newcommand{\DOWNML}[1]{\downarrow\!\!\{#1\}}

\newcommand{\VERTICAL}{\;  \mid\hspace{-3.0pt}\mid \; }
\newcommand{\UPML}[1]{\uparrow\!\!\{#1\}}
\newcommand{\EVAL}[1]{\mathsf{eval}(#1)}

\newcommand{\TYPEDEVAL}[2]{\mathsf{eval}^{#1}(#2)}

\newenvironment{GRAMMAR}{\[\begin{array}{rcl c rcl c rcl}}{\end{array}\]}
\newenvironment{RULES}{\[\begin{array}{c}}{\end{array}\]}
\newcommand{\CONV}{\Downarrow_{\lambda}}

\newcommand{\CONVCT}{\Downarrow_{\mathit{ct}}}
\newcommand{\CONVDT}{\Downarrow_{\mathit{dl}}}

\newcommand{\CONVMU}{\Downarrow_{\mathit{ul}}}

\newcommand{\LET}[3]{\mathsf{let}\ #1\ =\ #2\ \mathsf{in}\ #3}
\newcommand{\SUBST}[2]{[#1/#2]}
 \newenvironment{FIGURE}{\begin{figure}}{\end{figure}}
\newenvironment{TWOCOLUMNFIGURE}{\begin{figure*}}{\end{figure*}}
\newcommand{\CONST}[1]{\mathsf{#1}}
\newcommand{\AST}[1]{\mathsf{ast}_{\mathsf{#1}}}

\newcommand{\CONSTEVAL}{\CONST{eval}}
\newcommand{\CONSTTYPEDEVAL}[1]{\CONSTEVAL(#1)}
\newcommand{\CONSTAPPLY}{\CONST{promote}}
\newcommand{\CONSTVAR}{\CONST{var}}

\newcommand{\CONSTSTRING}{\CONST{string}}
\newcommand{\CONSTAPP}{\CONST{app}}
\newcommand{\CONSTLAM}{\CONST{lam}}

\newcommand{\CONSTINT}{\CONST{int}}
\newcommand{\CONSTADD}{\CONST{add}}
\newcommand{\CONSTMULT}{\CONST{mult}}

\newcommand{\CONSTREC}{\CONST{rec}}

\newcommand{\CONSTLIFT}{\CONST{lift}}
\newcommand{\TAG}[1]{{\mathsf{tag}}_{\mathsf{#1}}}

\newcommand{\TAGTYPE}[1]{\mathsf{Tag}_{#1}}

\newcommand{\TAGAPPLY}{\TAG{\CONSTAPPLY}}

\newcommand{\TAGINT}{\TAG{\CONSTINT}}

\newcommand{\ASTEVAL}[1]{\AST{\CONSTEVAL}(#1)}
\newcommand{\ASTTYPEDEVAL}[2]{\AST{\CONSTTYPEDEVAL{#1}}(#2)}
\newcommand{\ASTAPPLY}[1]{\AST{\CONSTAPPLY}(#1)}
\newcommand{\ASTVAR}[1]{\AST{\CONSTVAR}(#1)}

\newcommand{\ASTSTRING}[1]{\AST{\CONSTSTRING}(#1)}
\newcommand{\ASTAPP}[2]{\AST{\CONSTAPP}(#1, #2)}
\newcommand{\ASTLAM}[2]{\AST{\CONSTLAM}(#1, #2)}

\newcommand{\ASTINT}[1]{\AST{\CONSTINT}(#1)}
\newcommand{\ASTADD}[2]{\AST{\CONSTADD}(#1, #2)}
\newcommand{\ASTMULT}[2]{\AST{\CONSTMULT}(#1, #2)}

\newcommand{\ASTREC}[3]{\AST{\CONSTREC}(#1, #2, #3)}

\newcommand{\ASTLIFT}[1]{\AST{\CONSTLIFT}(#1)}

\newcommand{\IFTHEN}[2]{\mathsf{if}\; #1\;\mathsf{then}\; #2}
\newcommand{\IFTHENELSE}[3]{\IFTHEN{#1}{#2}\;\mathsf{else}\;#3}
\newcommand{\INT}{\mathsf{Int}}

\newcommand{\BOOL}{\mathsf{Bool}}
\newcommand{\FS}{\rightarrow}
\newcommand{\TYPES}[3]{#1 \vdash #2 : #3}

\newcommand{\STRING}{\mathsf{String}}

\newcommand{\POWER}{\text{\emph{power}}}

%% file: abstract.tex
\begin{abstract}
Homogeneous generative meta-programming (HGMP) enables the generation
of program fragments at compile-time or run-time. We present a
foundational calculus which can model both compile-time and run-time evaluated
HGMP, allowing us to model, for the first time, languages such as Template
Haskell. The calculus is designed such that it can be
gradually enhanced with the features needed to model many of the
advanced features of real languages. We demonstrate this
by showing how a simple, staged type system as found in Template Haskell can be added to the calculus.
\end{abstract}

%% file: introduction.tex
\section{Introduction}

\NI Homogeneous generative meta-programming (HGMP) enables program
fragments to be generated by a program as it is being either compiled
or executed. Lisp was the first language to support HGMP, and for many
years its only well known example. More recent languages such as
MetaML \cite{TahaW:mulstapitaa,TahaW:envClas} and Template
Haskell~\cite{SheardT:temmetpfh}
support varying kinds of HGMP. Simplifying slightly,
\emph{homogeneous} systems are those where the program that creates the
fragment is written in the same language as the fragment itself, in
contrast to heterogeneous systems such as the C preprocessor where two
different languages and/or systems are involved in the
generation~\cite{sheard__accomplishments_and_research_challenges_in_meta_programming,tratt__compile_time_meta_programming_in_a_dynamically_typed_oo_language}. Similarly, we use \emph{generative} to distinguish
HGMP from other forms of meta-programming such as reflection which focus on
analysing (but, in general, not changing) a system.

Perhaps surprisingly, given its long history, HGMP's semantics have
largely been defined by implementations~\cite{HermanD:thehygmac}. Some
aspects such as hygiene~\cite{adams15hygiene,HermanD:thehygmac} have
been studied in detail. There has also been extensive work on compile-time type-checked, run-time evaluated
HGMP, primarily  in the context of  MetaML and its
descendants~\cite{CalcagnoC:impmulsluagar,GanzS:macmulsc,InoueJ:reaabomsp,TahaW:mulstapitaa,TahaW:envClas}.

While we do not wish to advocate one style of HGMP over another,
we are not aware of work which provides a natural formal basis for the
style of HGMP found in languages such as Template Haskell and
Converge~\cite{tratt__compile_time_meta_programming_in_a_dynamically_typed_oo_language}
(broadly: compile-time evaluation of normal code with staged or dynamic type-checking).
Our intention in this paper is to directly model, without encodings, a wider range
of HGMP concepts than previously possible, in a simple yet expressive way,
facilitating greater understanding of how these concepts relate to one another.

The system we construct is based on a simple untyped $\lambda$-calculus, to
which we gradually add features and complexity, including a type
system. This allows us to model, for the first time, HGMP which
is evaluated at both compile-time (e.g.~Template Haskell-ish)
and run-time (e.g.~MetaML-ish) HGMP. As a side benefit, this also gives clear pointers for how
similar features can be added to `real' programming languages.

To summarise, this paper's key contributions are:

\begin{itemize*}
\item The first clear description of the design space of multiple languages and
confusingly similar, yet distinct, meta-programming systems.

\item The first calculus to be naturally to naturally model languages such as
Template Haskell.

\item The first calculus to be able to semi-systematically deal with
syntactically rich languages.

\item A demonstration of the calculus's simplicity by showing how it can be
easily extended to model (monotyped) systems such as Template Haskell's.
\end{itemize*}

%% The key contributions of this paper are:
%% \begin{itemize*}
%%     \item We present the first general calculus which models (untyped)
%%      compile-time and run-time HGMP as found in languages such as Template
%%      Haskell and MetaML.

%% \item We formalise type-safety for staged type-checking.

%% \item All (non-trivial) results have been defined and mechanically checked using
%%   Nominal Isabelle \cite{IsabelleFormalisation}.

%% \end{itemize*}

\section{HGMP design space}

Although HGMP can seem an easy topic to discuss, in reality its various flavours
and communities suffer greatly from incommensurability: important differences
are ignored; and similarities are obscured by terminology or culture. Since we
are not aware of previous work which tries to unify the various branches of the
HGMP family, it is vital that we start by sketching the major points in the
design space, so that we can be clear about both general concepts and the
specific terminology we use.

Figure~\ref{table_hgmpApproaches} summarises how some well known approaches sit
within this classification. We use `Lisp' as an over-arching term for a family
of related languages (from Common Lisp to Scheme) and `MetaML' to refer to MetaML 
and its descendants (e.g.~MetaOCaml). Similarly, we use
`JavaScript' to represent what are, from this paper's perspective, similar
languages (e.g.~Python and Ruby).

\input{table_hgmpApproaches}

\subsection{The HGMP subset of meta-programming}

The general area of meta-programming can be categorised in several different ways.
In this paper we consider homogeneous, generative meta-programming.

We use Sheard's definition of homogeneous and heterogeneous systems: 
``homogeneous systems [are those] where the meta-language and the object
language are the same, and heterogeneous systems [are those] where the
meta-language is different from the object-language''~\cite{sheard__accomplishments_and_research_challenges_in_meta_programming}.
The most
well known example of a heterogeneous generative meta-programming system is
C, where the preprocessor is both a separate system and language from
C itself. Heterogeneous systems are more flexible, but their power is
difficult to tame and reason
about~\cite{tratt__domain_specific_language_implementation_via_compile_time_meta_programming}.

In homogeneous systems in particular, we can then differentiate between
generative and reflective.  Reflection can introspect on run-time structures and
behaviour (as in e.g.~Smalltalk or
Self~\cite{bracha_ungra__mirrors_design_principles_for_meta_level_facilities_of_object_oriented_programming_languages}). In contrast, generative meta-programming explicitly constructs and
executes program fragments.

\subsection{Program fragment representation}

An important, yet subtle, choice HGMP languages must make is how to
represent the fragments that a program can generate. Three
non-exclusive options are used in practice: strings; abstract
syntax trees (ASTs); and upMLs (often called backquotes or
quasi-quotes). We now define each of these, considering their suitability
for our purposes.

\subsubsection{Strings}

In most cases, the simplest representation of a program is as a plain string.
Bigger strings can be built from smaller strings and eventually evaluated.
Evaluation typically occurs via a dedicated $\mathsf{eval}$ function at
run-time; a handful of languages provide a compile-time equivalent, which allows
arbitrary strings to be compiled into a source file. If such features are not
available, then a string can simply be saved to a temporary file, compiled, and
then run.

Representing program fragments as strings is trivial, terse, and can express
any program valid in the language's concrete syntax.\footnote{For most
languages this means that every possible program can be represented by
strings. A few languages forbid concrete syntax representations of valid ASTs
e.g.~Converge prevents variables beginning with \$ from being parsed,
as part of its hygiene system.} However, strings can express
nonsensical (e.g.~syntactically invalid) programs and prevent
certain properties (e.g.~hygiene or certain notions of type-safety) from being
enforced. Because of this, we believe that representing programs as
strings is too fragile to serve as a sound basis for a foundational model.

\subsubsection{ASTs}
\label{asts}

ASTs represent program fragments as a tree. For example, 
$2 + 3$ may be represented by the AST  $\ASTADD{\ASTINT{2}}{\ASTINT{3}}$.
ASTs are thus a simplification of a language's concrete syntax.
Exactly how the concrete syntax should be
simplified is influenced by an AST designer's personal tastes and preferences,
and different languages -- and, occasionally, different implementations of the
same language -- can take  different approaches. In general, ASTs are
designed to make post-parsing stages of a system easier to work with
(e.g.~type-checkers and code generators). HGMP languages which expose an
AST datatype also enable users directly to instantiate
new ASTs. Although ASTs generally allow semantically nonsensical programs to be
created (e.g.~referencing variables that are not defined), ASTs provide
fewer opportunities for representing ill-formed programs than strings.

By definition, every valid piece of concrete syntax must have a valid AST
representation (although not every valid AST may have a direct concrete syntax
representation; this possibility is rarely exploited, but see
e.g.~\cite{tratt__domain_specific_language_implementation_via_compile_time_meta_programming}
where it is used to help ensure lexical scoping for AST fragments generated in
module $M$ and inserted into $M'$).
STs are therefore the most fundamental representation of
programs and we use them as the basis of our calculus.

\subsection{UpMLs}
\label{upmls}

UpML (Up MetaLevel) is the name for the concept traditionally called
quasi-quote or backquote\footnote{Quasi-quotes were developed by Quine
  for working with logics~\cite{QuineWVO:mathlog}.}, which allow AST
(or AST-like) structures to be represented by quoted chunks of normal
program syntax\footnote{Quotation is typically indicated by syntactic
  annotations such as brackets, but it is also possible to eschew
  explicit markers altogether and use types to distinguish between
  programs and code as data \cite{JorringU:comstat,RompfT:ligmods}.}.
We have chosen the term `upML' to highlight an important relationship
with downMLs (see Section~\ref{language:CTHGMP}). UpMLs are often
used because they enable a familiar means of representing code. They
can also be used statically to guarantee various properties.

There are two distinct styles of upMLs in HGMP languages, which we now
discuss.  The most common style of upMLs is found in languages such as
Lisp and Template Haskell, where they are used as a `front-end' for
creating ASTs. While ASTs in such languages are powerful, even small
syntax fragments lead to deeply nested, unwieldy, ASTs. UpMLs allow
AST fragments to be directly built from a concrete syntax fragment
e.g.~the upML expression
\[
\UPML{2 + 3}
\]
evaluates to the AST
\[
\ASTADD{\ASTINT{2}}{\ASTINT{3}}.
\]
UpMLs can contain holes which are expressed using a downML
$\DOWNML{\ldots}$. The expression in the hole is expected to evaluate
to an AST. For example, if the function $f$ returns the AST equivalent
of $2 + 3$ -- in other words, $f$ returns $\ASTADD{\ASTINT{2}}{\ASTINT{3}}$ -- then
\[
\UPML{\DOWNML{f\;()}\ * 4}
\]
will have an intermediate evaluation equivalent to $\UPML{(2 + 3) *
  4}$, leading to the eventual AST
\[
\ASTMULT{\ASTADD{\ASTINT{2}}{\ASTINT{3}}}{\ASTINT{4}}.
\]
In this model, ASTs are the fundamental construct and upMLs a
convenience.

In their less common style -- found only, to the best of our
knowledge, in MetaML and its descendants -- upMLs are a datatype in
and of their own right, and do not represent ASTs. This
has the shortcoming that it cannot represent some reasonable
forms of meta-programming.  For example, as
discussed in~\cite{SheardT:temmetpfh}, one can not use this form
of upMLs to create projection functions such as:
\[
   \qquad
   (n, i)
         \quad \mapsto \quad
   \text{code of}\ \lambda (x_0, ..., x_{n-1}). x_i
\]
In contrast, one can always use (perhaps laboriously) AST constructors
to create such functions. A related problem relates to the tight
coupling of upMLs and a language's concrete grammar which may not
allow sub-constructs (e.g.~an \texttt{else} clause) to be used
in isolation and/or the location of holes can be ambiguous (if a hole
follows an $\mathsf{if}$ construct, is it expected to be filled with
an $\mathsf{else}$ clause, or a separate expression that follows, but is not
part of, the $\mathsf{if}$?).

Our experience is that languages without upMLs are prohibitively
difficult to use, and find little traction with users.  We therefore
add upMLs as an optional extension to our calculus. We choose upMLs as
a 'front-end’ for creating ASTs, due to the greater expressivity and
ubiquity of this approach.

\subsection{Compile-time vs.~run-time execution}

In order to be useful, HGMP program fragments must at some point be
run. Broadly speaking, execution happens at either
compile-time (when the wider program is being compiled but not executed) or
run-time (as part of normal program execution). Different languages allow
evaluation at compile-time and/or run-time depending on the language: Template
Haskell can evaluate ASTs only at compile-time; JavaScript can evaluate strings
only at run-time; while Lisp can evaluate ASTs and strings at both
compile-time and run-time.

Run-time evaluation is conceptually simple: a normal user-level function,
conventionally called $\mathsf{eval}$, takes in a program fragment (as a string or
AST, depending on the language) and evaluates it. Every time the program is
run, $\mathsf{eval}$ is called anew, as any other user-level function.

Compile-time evaluation is trickier and is represented in our approach with
top-level downMLs $\DOWNML{\ldots}$ (i.e.~a downML that is not nested inside an upML).
When a top-level downML is encountered, the code inside it is evaluated
before the surrounding program; that code must evaluate to an AST which
then overwrites the downML before normal compilation resumes. In other words,
all top-level downMLs are evaluated and `eliminated' before run-time execution.
Once the top-level downMLs are evaluated, a new program is constructed. No
matter how many times that new program is evaluated, the top-level downMLs are not
-- cannot be! -- reevaluated. We sometimes say that the effects of compile-time evaluation are 'frozen'
in the resulting program.

The practical effects of run-time and compile-time evaluation are
best seen by example. For example,
the following program evaluates code at run-time using $\mathsf{eval}$ 
and prints \texttt{1 3 6} (where
`\texttt{;}' is the sequencing operator):
\begin{lstlisting}
   print(1); 
   print(2 + eval(print(3); ASTInt(4)))
\end{lstlisting}
Replacing the $\mathsf{eval}$ with a downML
leads to a program which prints \texttt{3 1 6}:
\begin{lstlisting}
   print(1); 
   print(2 + |$\DOWNML{\texttt{print(3); } \texttt{ASTInt(4)}}$|)
\end{lstlisting}
These two possibilities have various implications. For example, compile-time
evaluation allows ASTs to interact with early stages of the programming
language's semantics and can introduce new variables into scope.
In contrast, run-time evaluation can reference variables
but not change those in scope in any way. There is also a significant
performance difference: if a calculation can be moved from run-time to
compile-time, it then has no run-time impact. 

\subsection{Implicit and explicit HGMP}

Compile-time evaluation can also be subdivided into explicit and implicit
flavours.  In Lisp, `macros' are special constructs
explicitly identified by a user; `function calls' whose name reference a
macro are identified by the compiler and the macro evaluated at compile-time
with the `function call' arguments passed to it. Since one cannot tell by
looking at a Lisp function call in isolation whether its arguments will be
evaluated at compile-time or run-time, it allows Lisp programmers to `extend'
the language transparently to the user. Although implicit evaluation has
traditionally been seen as more problematic in syntactically rich languages,
Honu~\cite{rafkind12honu} shows that a Lisp-like macro system can be embedded in
such languages. However, languages such as Template Haskell take a different
approach, explicitly identifying the locations where compile-time evaluation
will happen using downMLs, allowing the user to call arbitrary user code.
To an extent, the difference between implicit and explicit HGMP is cultural and
without wishing to
pick sides, in this paper we concentrate on explicit HGMP. This allows us to
concentrate on the fundamentals of HGMP without the parsing considerations that
are generally part of implicit HGMP.

\subsection{HGMP vs.~macro expansion}
\label{hgmp vs macro expansion}

Systems such as Template Haskell share the same compile-time and
run-time language (with the small exception that the run-time language does not
feature upMLs): the compile-time evaluation of code uses the same evaluation
rules as run-time evaluation. While some Lisp systems share this model, some do
not. Most notably Scheme, and its descendent Racket, use a macro
expander~\cite{flatt12macros}.  Rather than evaluate arbitrary Lisp code, Scheme
and Racket macros share the same syntax as, but a different evaluation semantics
to, their surrounding language. Thus, in our terminology, \texttt{define-syntax}
is not a HGMP system. Modern Scheme and Racket systems have an additional macro
system \texttt{syntax-case} which does allow HGMP.

%% file: table_hgmpApproaches.tex
\begin{figure*}[tb]
\centering
\begin{tabular}{lccccc}
\toprule
Language                     & Strings   & ASTs      & UpMLs       & Compile-time HGMP & Run-time HGMP \\
\midrule
Converge                     & $\bullet$ & $\bullet$ & $\bullet$    & $\bullet$         & $\bullet$ \\
JavaScript                   & $\bullet$ & $\circ$   & $\circ$      & $\circ$           & $\bullet$ \\
Lisp                         & $\bullet$ & $\bullet$ & $\bullet$    & $\bullet$         & $\bullet$ \\
MetaML                       & $\circ$   & $\circ$   & $\bullet$    & $\circ$           & $\bullet$ \\
Template Haskell             & $\circ$   & $\bullet$ & $\bullet$    & $\bullet$         & $\circ$ \\
Scala (\texttt{scala.meta})  & $\circ$   & $\bullet$ & $\bullet$    & $\bullet$         & $\bullet$ \\
\bottomrule
\end{tabular}
\caption{A high-level characterisation of various HGMP languages.}
\label{table_hgmpApproaches}
\end{figure*}

%% file: language.tex
\section{A simple HGMP calculus}
\label{calculus}

In this section, we define the minimal calculus which does interesting
HGMP so that we can focus on the core features. In later sections
we enrich this calculus with more advanced constructs. Our starting point is a standard
call-by-value (CBV) $\lambda$-calculus whose grammar is as follows:
\begin{GRAMMAR}
  M 
     &::=&
  x
     \VERTICAL
  MN
     \VERTICAL
  \lambda x.M
     \VERTICAL
  c
     \VERTICAL
  M + N 
     \VERTICAL
  ...  
\end{GRAMMAR}
Here, $x$ ranges over variables and $c$ over constants (e.g.~strings, integers).
We include $+$ as an example of a wide class of common
syntactic constructs.

\subsection{ASTs}
\label{sec:simple asts}

As discussed in Section~\ref{asts}, ASTs are the most fundamental form
of representing programs in HGMP. In essence, every element of the
calculus must have a representation as an AST. Of the syntactic
constructs in the $\lambda$-calculus, constants, applications, and additions are most easily
modelled; both variables and $\lambda$-abstractions require the
representation of variables. We model variables as strings, which makes
modelling later HGMP features easier and matches `real' systems.
We thus extend the $\lambda$-calculus as follows:
\begin{GRAMMAR}
  M
     &::=&
  ...
  \VERTICAL
     \AST{t}(\VEC{M})
     \\[2mm]
  t
     &::=&
     \CONSTVAR
  \VERTICAL
     \CONSTAPP
  \VERTICAL
     \CONSTLAM
  \VERTICAL
     \CONSTINT
  \VERTICAL
     \CONSTSTRING
  \VERTICAL
     \CONSTADD
  \VERTICAL
     ...  
\end{GRAMMAR}

\NI We write $\VEC{M}$ for tuples $(M_1, ..., M_n)$ with $|\VEC{M}|$ denoting the length of the tuple.
An AST constructor $\AST{t}(\VEC{M})$ takes $|M| + 1$ arguments.  The first
argument $t$ is a \emph{tag}  which specifies the specific AST datatype, and the rest
of the arguments are then relative to that datatype. For example $\ASTVAR{"x"}$ is
the AST representation of the variable $x$, $\ASTLAM{\ASTSTRING{"x"}}{M}$ is
the AST representation of $\lambda x.N$, and $\ASTINT{3}$ is the AST representation of the constant $3$.

\subsection{Compile-time HGMP}\label{language:CTHGMP}

To model compile-time HGMP, we extend the calculus with a new construct downML,
which provides a way of syntactically defining the points in a program where
compile-time HGMP should occur (we do not need a tag for downMLs
for reasons that will become clear later):
\begin{GRAMMAR}
  M &::=&  ... \VERTICAL \DOWNML{M}
     &\quad\quad
  t & ::= &  ...
\end{GRAMMAR}
In essence, a downML $\DOWNML{M}$ is an
expression which must be evaluated at compile-time, i.e.~before the
rest of the program is executed.  To model this, we find ourselves in
the most complex and surprising part of our calculus: we have distinct but interacting
reduction relations for the compile-time and run-time stages.

\input{figure_reductions_simple}

Figure~\ref{figure_reductions_noEval} shows the reduction rules for our simple
system. We use  big-step semantics for brevity. There are three reduction relations:
\begin{description}
    \item[$\CONVCT$] models a compiler. It takes a program,
        possibly containing downMLs, as input and produces a 
        program with no downMLs as output. It does this by recursively scanning
        through the input program looking for downMLs and evaluating them.
        Normal $\lambda$-calculus terms are copied from input to output
        unchanged. When a downML is encountered, the rule [\RULENAME{DownML ct}], explained below,
        evaluates the expression inside the downML. Assuming that expression
        returns an AST, the $\CONVDT$ relation turns the AST
        into a normal program which then overwrites the
        downML. For example, $(\lambda z.z)\DOWNML{\ASTSTRING{(\lambda y. y) "x"}} \CONVCT (\lambda z.z)"x"$.

    \item[$\CONVDT$] models the conversion of ASTs into `normal' programs.
        In our case, this means converting ASTs into programs
        (e.g.~$\ASTSTRING{"x"} \CONVDT "x"$).  As this may suggest, $\CONVDT$ is
        a simple relation which can be semi-mechanically created from the AST
        structure of the language.
    
    \item[$\CONV$] models run-time execution. The rules are normal 
        $\lambda$-calculus CBV reduction rules augmented with the minimum rules to
        evaluate ASTs.
\end{description}
Put another way, $\CONVCT$ and $\CONV$ are the key reduction relations,
which allow us to accurately capture the reality that a program is
compiled once but run many times:
\[
	\underbrace{M \CONVCT}_{\text{compile-time}} A\ 
	\underbrace{\CONV V}_{\text{run-time}}
\]

The key rule in $\CONVCT$ is [\RULENAME{DownML ct}]. Its
left-most premise $M \CONVCT A$ first recursively scans for downMLs nested in
$M$. The middle premise $A \CONV B$ is the heart of the rule, evaluating the
expression to produce an AST using normal $\lambda$-calculus evaluation.
The simplicity of this premise belies its importance:  the expression $A$ 
can perform arbitrary computation.
For the time being, we assume that the expression
returns an AST (we defer consideration of erroneous computations to
Section~\ref{stuck programs}). That AST is then converted into a normal
program which overwrites the downML.  The program
resulting from the $\CONVCT$ relation can then be run as many times
with the $\CONV$ relation as desired.  
Figure~\ref{table_exampleReduction1} shows a fully worked-out example
of a compile-time program and the $\CONVCT$ relation. 

Those familiar with the innards of systems such as Template Haskell
will notice that our system is able to simplify their workings. Whereas a `real'
compiler first has to convert a string program to a parse tree and then produce
ASTs from it, the $\lambda$-calculus naturally fulfils this role. We are thus
saved from having to introduce a fourth reduction relation.

\subsubsection{Scoping}

Our simple calculus intentionally allows variables to be captured dynamically.
Although this naturally follows from the reduction rules, we now explicitly
explain how this occurs and why. First, we note that the [\RULENAME{App}]
rule in the definition of $\CONV$ uses traditional capture-avoiding substitution
$M\SUBST{N}{x}$. Note that we do not need to extend the definition to downMLs,
which will have been removed by $\CONVCT$ before substitution is applied.

We can create AST variables containing arbitrary variables (e.g.~$\ASTVAR{"x"}$)
which downMLs will then turn into normal programs. Consider the following two
programs and their compilation:
\begin{itemize}
\item 
  $\lambda x. \DOWNML{ \ASTVAR{"x"}}
     \CONVCT
  \lambda x.x$.

\item 
  $\lambda y. \DOWNML{ \ASTVAR{"x"}}
     \CONVCT
  \lambda y.x$.
\end{itemize}
\noindent As these examples suggest, our calculus is not hygienic, and thus
allows variables to be captured. This is a deliberate design decision for two
reasons. First, not all languages that we wish to model have an explicit notion
of hygiene, instead providing a function which generates fresh (i.e.~unique)
names (conventionally called $\mathsf{gensym}$). Second, there is not, as yet, a
single foundational style of hygiene, and different languages take subtly
different approaches. We talk about  possible
future directions for hygiene in Section~\ref{conclusions}.

\subsection{Run-time HGMP}
\label{sec:runtimehgmp}

Having introduced compile-time HGMP, we now have all the basic tools needed to
introduce run-time HGMP. We follow the Lisp tradition and use a function called
$\CONSTEVAL$. Unlike downMLs, $\CONSTEVAL$s are not eliminated at
compile-time: they are, in essence, normal $\lambda$-calculus functions. We extend
the calculus grammar (including an AST equivalent) as follows:
\[
  M ::=
     ...
  \VERTICAL
     \EVAL{M}
  \qquad\qquad
  t 
     ::=
  ...
     \VERTICAL
  \CONSTEVAL
\]

\input{figure_reductions_eval}

\NI The additional reduction rules for $\CONSTEVAL$ are shown in
Figure~\ref{fig:evalrules}.  [\RULENAME{Eval rt}] reduces  $M$ to a value, which
must be an AST, and which is then turned into a normal $\lambda$ term and
executed. Note that unlike compile-time HGMP, $\CONSTEVAL$ cannot introduce new variables into a scope.  A
detailed example of running $\CONSTEVAL$ is given in Figure
\ref{table_exampleReduction2}.

\input{table_exampleReduction1}

\section{Enriching the calculus}
\label{sec:enriched calculus}

The simple calculus of the previous section allows readers to concentrate on the
core of our approach. However, it is too spartan to model
important properties of real languages. In this section, we enrich the simple
calculus with further features which add complexity but allow us to model real
languages.

\subsection{Higher-order ASTs}
\label{higherorderasts}

\input{figure_reductions_promote}

The simple system in Section~\ref{sec:simple asts} does 
not allow higher-order meta-programming (e.g.~meta-meta-programming).
While the simple ASTs we introduced in Section~\ref{sec:simple asts} are
sufficient to represent normal $\lambda$-calculus terms as an AST, 
programs with ASTs cannot be represented as ASTs. 
While not all real languages (e.g.~Template
Haskell) allow higher-order meta-programming, many do (e.g.~MetaML and
Converge). We thus introduce higher-order ASTs now to make
the presentation of later features consistent.

Higher-order ASTs need a means to represent programs that
can create ASTs as ASTs themselves. There are many plausible
ways that this could be done: the mechanism we settled upon allows
extra syntactic elements to be easily added by further extensions. The
basis of our approach is a new datatype $\AST{\CONSTAPPLY}(M,
\VEC{N})$ which allows an arbitrary AST with a tag $M$ and parameters $\VEC{N}$
to be promoted up a meta-level. We thus need to introduce a way for
programs to reference tags arbitrarily, and extend the syntax as follows:
\[
  M
     ::=
     ...
  \VERTICAL
     \TAG{\mathrm{t}}
  \qquad\qquad
  t
     ::=
  ...
     \VERTICAL
  \CONSTAPPLY
  \]
\noindent The corresponding reduction rules are in Figure
\ref{figure_reductions_promote}. Note that tags are normal values in the
calculus so that one can write programs which can create 
higher-order ASTs. Promoted ASTs can then be reduced one meta-level with the
existing $\CONVDT$ relation. For example, $\AST{\CONSTAPPLY}(\textsf{string},
\AST{\CONSTSTRING}("x")) \CONVDT \AST{\CONSTSTRING}("x")$.

\subsection{UpMLs}
\label{sec:upmls}

\input{figure_reductions_mu}

Using AST constructors alone to perform HGMP is tiresome---while it gives
complete flexibility, the sheer verbosity of such an approach quickly
overwhelms even the most skilled and diligent programmer. UpMLs ameliorate
this problem by allowing concrete syntax to be used to represent ASTs (see
Section~\ref{upmls}). Since, depending on a language's syntax, upMLs can be less
expressive than ASTs, we model upMLs as a
transparent compile-time expansion to the equivalent AST constructor calls
e.g.~$\UPML{2} \CONVCT \AST{\CONSTINT}(2)$.

To add UpMLs to our language, we first extend the grammar as follows:
\begin{GRAMMAR}
  M 
     &::=&
  ... 
     \VERTICAL
  \UPML{M}
     &\quad\quad
  t & ::= &  ...
\end{GRAMMAR}
\NI Note that, like downMLs, upMLs have disappeared after the  compile-time
stage, so we have no need to make an AST equivalent of them.

Figure~\ref{figure_reduction:mu} shows the reduction rules needed for upMLs
including the new $\CONVMU$ reduction relation which handles the upML to
AST conversion. When, during the recursive sweep of a program by the
$\CONVCT$ reduction relation, an upML is encountered, it is handed over to the
$\CONVMU$ reduction relation which translates a $\lambda$-term into its
AST equivalent.

The major subtlety in the new rules relates to an important practical need.
UpMLs on their own can only construct ASTs of a fixed `shape' and are thus
rather limited. Languages with upMLs (or their equivalents) therefore allow holes to
be put into them where arbitrary ASTs can be inserted; in essence, the upML
serves as a template with defined points of variability. In some languages
(e.g.~Converge) holes inside upMLs are syntactically differentiated from holes
outside, but we use
downMLs to represent such `inner' holes. In the same way as top-level downMLs, inner downMLs are expected
to return an AST; unlike top-level downMLs,
they are evaluated at run-time not compile-time. The [\RULENAME{DownML ul}] rule 
therefore simply runs the expression inside it through the $\CONVCT$
reduction relation and uses the result as-is. This allows examples such as the
following:
\[
   \UPML{2 + \DOWNML{\UPML{3 + 4}}} 
      \quad\CONVCT\quad
   \ASTADD{\ASTINT {2}}{\ASTADD{\ASTINT {3}}{\ASTINT 4} }
\] 
\noindent In our model, upMLs are simple conveniences for AST
construction, rather as they were in early Lisp implementations. More
recent languages (e.g.~Scheme, Template Haskell) use upMLs
in addition as a means of ensuring referential transparency and
hygiene~\cite{ClingerW:mactw}. Our formulation of upMLs is designed to
open the door for such possibilities, but it is beyond the scope of this
paper to tackle them.

\subsubsection{The relationship between compile-time levels}
\label{meta-levels}

Readers may wonder why we have chosen the names upML and
downML for what are often called backquote / quasi-quote and macro call /
splice respectively. We build upon an observation from
MetaLua~\cite{fleutot_tratt__contrasting_compile_time_meta_programming_in_metalua_and_converge}
that these two operators are more deeply connected than often considered, though
our explanation is somewhat different. Our starting point is to note that,
during compilation, there are three stages that a compiler can go through:
normal compilation; converting upMLs to ASTs; and running user code in a downML.
UpMLs / downMLs not only control which stage the compiler is in at any point during compilation,
but have a fundamental relation to ASTs which we now investigate.

We call the normal compilation stage level 0. AST constructors in normal $\lambda$-terms
are simply normal datatype constructors. When we
encounter a top-level upML, we shift stage `up' to level +1. In this level we take code and
convert it into an AST which represents the code. When we encounter a top-level
downML, we shift stage `down' to level -1. In this level we take code and run it.

The basic insight is that the compiler level corresponds to the ASTs created or
consumed: at level 0 we neither create or consume ASTs; at level 1 we create
them (with upMLs); and at level -1, we consume ASTs (downMLs must evaluate to an
AST, which is then converted to a normal $\lambda$-term).

Building upon this, we can see that this notion naturally handles downMLs nested
within upMLs (and vice versa), bringing out the symmetry between the two
operators, which can cancel each other out. Consider a program which
nests a downML in an upML (i.e.~$\UPML{\DOWNML{M}}$). How is the program $M$
dealt with? Compilation starts at level 0; the upML shifts it to level 1; and
the downML shifts it back to level 0. Thus we can see that $M$ is handled at the
normal compilation level and neither creates or consumes ASTs at compile-time
(the fact that, at run-time, $M$ ultimately needs to evaluate to an AST is
irrelevant from a compile-time perspective). Similarly, consider an upML nested
inside a downML (i.e.~$\DOWNML{\UPML{M}}$). Since the downML shifts the compiler
to level -1 and the upML shifts it back to level 0, we can see that the end effect
is that $M$ is dealt with as if it had always been at the normal compilation level.

In fact, the notion of these 3 levels (-1, 0, +1) is sufficient to explain
arbitrarily nested downMLs and upMLs. For example, we can clearly see that two
nested upMLs (i.e.~$\UPML{\UPML{M}}$) create an AST representation of $M$ (at
level 2) which can be turned back into a normal $\lambda$-term by two
nested downMLs (operating at level -2). As this suggests, unlike systems such as
MetaML, we do not need to label expressions as belonging to a certain level, nor
do we need to do anything special to handle levels extending to $-\infty$ or
$+\infty$.

\subsection{Lifting}\label{Lifting}

Most HGMP languages allow semi-arbitrary run-time values to be lifted up a meta-level
(e.g.~an integer 3 to be converted to an AST $\ASTINT{3}$). In some
languages lifting is implicit (e.g.~in Template Haskell, a variable inside an
upML which references a definition outside the upML, and which is of a simple
type such as integers, is implicitly lifted), while in others it is explicit (e.g.~Converge forces all
lifting to be explicit). All  the languages we are aware of that use implicit
lifting determine this statically, and can be trivially translated to
explicit lifting. We thus choose to model explicit lifting.
We extend the grammar as follows:
\begin{GRAMMAR}
   M &::=& ... \VERTICAL \LIFT{M} 
      &\qquad
   t &::=& ... \VERTICAL \CONSTLIFT 
\end{GRAMMAR}
\noindent Figure~\ref{figure_reductions_lift} shows the additional reduction
rules. The rules for the $\CONVCT$, $\CONVDT$, and $\CONVMU$ relations are
mechanical. Capture-avoiding substitution is given as $\LIFT{M}\SUBST{N}{x} = 
\LIFT{M\SUBST{N}{x}}$. The rules for $\CONV$ show that $\mathsf{lift}$ is a polymorphic
function, turning values of type $T$ into an AST $\textsf{ast}_{\textsf{T}}$
e.g.~$\textsf{lift}(2 + 3) \CONV \ASTINT{5}$.

The relation between upMLs and $\textsf{lift}$ can be seen from the following
examples (where $\circ$ represents relational composition):
\begin{itemize}

\item $\UPML{2+3} \CONVCT \ASTADD{\ASTINT{2}}{\ASTINT{3}}$
\item $\UPML{2+3} \;(\CONVCT \circ \CONV)\; \ASTADD{\ASTINT{2}}{\ASTINT{3}}$
\item $\LIFT{2+3} \CONVCT \LIFT{2+3}$
\item $\LIFT{2+3} \;(\CONVCT \circ \CONV)\; \ASTINT{5}$

\end{itemize}

\subsection{Cross-level variable scoping}

\input{figure_substitution_letdownmls}

The downMLs modelled in Section~\ref{language:CTHGMP} run each
expression in a fresh environment with no link to the outside
world. While in theory this is sufficiently expressive, in practice it is
restrictive: downMLs cannot share code, and so each downML must include
within it a copy of every library function it wishes to use.  Languages
such as Template Haskell therefore allow variables defined outside downMLs
(e.g.~functions) to be referenced within a downML. Different languages have
subtly different mechanisms to define which variables are available within a downML
(e.g.~Converge allows, with some restrictions, variables defined within a module $M$
to be used in a downML within that module; Template Haskell only
allows variables imported from other modules to be used in a downML), and we do not
wish to model the specifics of any one language's scheme.

We therefore provide a simple abstraction which can be used to model
the scoping rules of different languages. The \emph{letdownML}
construct $\LETCT{x}{M}{N}$ makes a program $M$ available as $x$ to
$N$ at compile-time (i.e.~including inside downMLs). We
extend the grammar as follows:
\begin{GRAMMAR}
   M &::=& ... \VERTICAL \LETCT{x}{M}{N}
      &\qquad
   t &::=& ...
\end{GRAMMAR}
\NI The additional reduction rule for letdownML is given in
Figure~\ref{substitution_letdownmls}. As this shows, letdownMLs are let bindings
that are performed at compile-time rather than run-time.
Figure~\ref{substitution_letdownmls} therefore also defines the additional
substitution rules required.

\subsection{Examples}

The staged power function~\cite{CzarneckiK:dslimpimthac} has become a
standard way of comparing HGMP approaches. The idea is to specialise
the function $\lambda nx.x^n$ with respect to its first argument.  This is more
efficient than implementations with variable exponent, provided the
cost of specialisation is amortised through repeated use at run-time.  To model
this in our calculus, we assume the existence of the standard recursion operator
$\mu g.\lambda x. M$  that makes $g$ available for recursive calls in
$M$. The staged power function then becomes:
\begin{align*}
  M
     &=
  \mu p. 
     \lambda n. 
     \IFTHENELSE
         {n = 1}
         {\!\!\UPML{x}}
         {\!\!\UPML{x\, \times\! \DOWNML{p\ (n-1)}}} \\
  power
     &=
  \lambda n.\UPML{\lambda x.\DOWNML{M\ n}}
\end{align*}
For example $power\ 3$ reduces to an AST equivalent to that generated by
$\UPML{\lambda x . x \times x \times x}$.  The function $power$ can be used to
specialise code at compile-time:
\[
   \LET{cube}{\DOWNML{\POWER\ 3}}{(cube\ 4) + (cube\ 5)}
\]
and at run-time:
\[
   \LET{cube}{\EVAL{\POWER\ 3}}{(cube\ 4) + (cube\ 5)}
\]
By stretching the example somewhat, we can also show higher-order HGMP in
action. Assume we wish to produce a variant of \emph{power} which takes
one part of the exponent early on in a calculation, with the second
part known only later (e.g.~because we want to compute $\lambda n.x^{m+n}$ frequently for a small number
of different $n$ that become available after $m$ is known). We can then use the following
higher-order meta-program for this purpose:
\[
    power_{ho}
     =
  \lambda m.\UPML{\lambda n. \UPML{\lambda x.\DOWNML{M\ (m+n)}}}
\]
and use it in a number of different ways e.g.:
\[
    \LET{cube}{\DOWNML{\DOWNML{power_{ho}\ 1}\ 2}}{cube\ 4}
\]
which specialises both arguments at compile-time, or
\[
    \LET{f}
    {\DOWNML{power_{ho}\ 1}}
      \LET{cube}{f\ 2}{\EVAL{cube}\ 4}
\]
where the first argument is specialised at compile-time, and the
second at run-time.

\input{figure_reductions_lift}

%% file: figure_reductions_simple.tex
\begin{TWOCOLUMNFIGURE}
\begin{RULES}
  \ZEROPREMISERULENAMEDRIGHT
  {
    x \CONVCT x
  }{Var ct}
  \quad
  \TWOPREMISERULENAMEDRIGHT
  {
    M \CONVCT A
  }
  {
    N \CONVCT B
  }
  {
    MN \CONVCT AB
  }{App ct}
  \quad
  \ONEPREMISERULENAMEDRIGHT
  {
    M \CONVCT A
  }
  {
    \lambda x.M \CONVCT \lambda x.A
  }{Lam ct}
  \quad
  \ZEROPREMISERULENAMEDRIGHT
  {
    c \CONVCT c
  }{Const ct}
  \\\\
  \TWOPREMISERULENAMEDRIGHT
  {
    M \CONVCT A
  }
  {
    N \CONVCT B
  }
  {
    M + N \CONVCT A+B
  }{Add ct}
  \quad
  \ONEPREMISERULENAMEDRIGHT
  {
    M_i \CONVCT A_i
  }
  {
    \AST{\mathrm{t}}(\VEC{M}) 
       \CONVCT
    \AST{\mathrm{t}}(\VEC{A}) 
  }{Ast$_c$ ct}

  \\\\
  \THREEPREMISERULENAMEDRIGHT
  {
    M \CONVCT A
  }
  {
    A \CONV B
  }
  {
    B \CONVDT C
  }
  {
    \DOWNML{M} \CONVCT C
  }{DownML ct}
\\\\
\hdashline[0.5pt/2pt]
\\
  \ZEROPREMISERULENAMEDRIGHT
  {
    \ASTVAR{"x"} \CONVDT x
  }{Var dl}
  \quad
  \TWOPREMISERULENAMEDRIGHT
  {
    M \CONVDT M'
  }
  {
    N \CONVDT N'
  }
  {
    \ASTAPP{M}{N} \CONVDT M'N'
  }{App dl}
  \\\\
  \TWOPREMISERULENAMEDRIGHT
  {
    M \CONVDT "x"
  }
  {
    N \CONVDT N'
  }
  {
    \ASTLAM{M}{N} \CONVDT \lambda x.N'
  }{Lam dl}
  \quad
  \ZEROPREMISERULENAMEDRIGHT
  {
    \ASTINT{n} \CONVDT n
  }{Int dl}
  \\\\
  \ZEROPREMISERULENAMEDRIGHT
  {
    \ASTSTRING{"x"} \CONVDT "x"
  }{String dl}
  \quad
  \TWOPREMISERULENAMEDRIGHT
  {
    M \CONVDT M'
  }
  {
    N \CONVDT N'
  }
  {
    \ASTADD{M}{N} \CONVDT M' + N'
  }{Add dl}
\\\\
\hdashline[0.5pt/2pt]
  \\
  \ZEROPREMISERULENAMEDRIGHT
  {
    \lambda x.M \CONV \lambda x.M
  }{Lam}
  \quad
  \THREEPREMISERULENAMEDRIGHT
  {
    M \CONV \lambda x.M'
  }
  {
    N \CONV N'
  }
  {
    M'\SUBST{N'}{x} \CONV L
  }
  {
    MN \CONV L
  }{App}
  \\\\
  \THREEPREMISERULENAMEDRIGHT
  {
    ...
  }
  {
    M_i \CONV N_i
  }
  {
    ...
  }
  {
    \AST{\mathrm{t}}(\VEC{M}) 
       \CONV
    \AST{\mathrm{t}}(\VEC{N}) 
  }{Ast$_c$}
\end{RULES}

\caption{Key big-step reduction rules for the CBV semantics of our
  simple calculus. Some standard rules (e.g.~$\CONV$ for addition) are
  omitted for brevity. }\label{figure_reductions_noEval}

\end{TWOCOLUMNFIGURE}

%% file: figure_reductions_eval.tex
\begin{figure}
\begin{RULES}
  \THREEPREMISERULENAMEDRIGHT
  {
    L \CONV M
  }
  {
    M \CONVDT N
  }
  {
    N \CONV N'
  }
  {
    \EVAL{L} \CONV N'
  }{Eval rt}
  \quad
  \ONEPREMISERULENAMEDRIGHT
  {
    M \CONVCT N
  }
  {
    \EVAL{M} \CONVCT \EVAL{N}
  }{Eval ct}
  \quad
  \ONEPREMISERULENAMEDRIGHT
  {
    M \CONVDT N
  }
  {
    \ASTEVAL{M} \CONVDT \EVAL {N}
  }{Eval dl}
\end{RULES}
\caption{Additional reduction rules for $\CONSTEVAL$.}
\label{fig:evalrules}
\end{figure}

%% file: table_exampleReduction1.tex
\begin{figure*}[tb]

\centering
\begin{prooftree}
  \AXIOM{x \CONVCT x}{}
  \UNARYINFERENCE{\lambda x.x \CONVCT \lambda x.x}{}
  \AXIOM{x \CONVCT x}{}
  \UNARYINFERENCE{\lambda x.x \CONVCT \lambda x.x}{}
  \AXIOM{7 \CONVCT 7}{}
  \UNARYINFERENCE{\ASTINT{7} \CONVCT \ASTINT{7}}{}
  \BINARYINFERENCE{(\lambda x.x)\ASTINT{7} \CONVCT (\lambda x.x)\ASTINT{7}}{}
  \AXIOM{...}{}
  \UNARYINFERENCE{(\lambda x.x)\ASTINT{7} \CONV \ASTINT{7}}{}
  \AXIOM{\ASTINT{7} \CONVDT 7}{}
  \TRINARYINFERENCE{\DOWNML{(\lambda x.x)\ASTINT{7}} \CONVCT 7}{}
  \BINARYINFERENCE{(\lambda x.x) \DOWNML{(\lambda x.x)\ASTINT{7}} \CONVCT (\lambda x.x)7}{}
\end{prooftree}

\vspace{4mm}

\begin{prooftree}
  \AXIOM{\lambda x.x \CONV \lambda x.x}{}
  \AXIOM{(\lambda x.x)\ASTINT{7} \CONV \ASTINT{7}}{}
  \AXIOM{\ASTINT{7} \CONVDT 7}{}
  \AXIOM{7 \CONV 7}{}
  \TRINARYINFERENCE{\EVAL{(\lambda x.x)\ASTINT{7}} \CONV 7}{}
  \AXIOM{x\SUBST{7}{x} \CONV 7}{}
  \TRINARYINFERENCE{(\lambda x.x)(\EVAL{(\lambda x.x)\ASTINT{7})} \CONV 7}{}
\end{prooftree}

\caption{Examples of compile-time HGMP (top) and run-time HGMP (bottom).}
\label{table_exampleReduction1}
\label{table_exampleReduction2}
\end{figure*}

%% file: figure_reductions_promote.tex
\begin{FIGURE}
\begin{RULES}
  \THREEPREMISERULENAMEDRIGHT
  {
    ...
  }
  {
    M_i \CONVCT A_i
  }
  {
    ...
  }
  {
    \ASTAPPLY{\TAG{t}, \VEC{M}} \CONVCT \ASTAPPLY{\TAG{t}, \VEC{A}}
  }{Promote ct}
  \\\\
  \FIVEPREMISERULENAMEDRIGHT
  {
    L \CONVDT \TAG{t}
  }
  {
    t \neq \CONSTAPPLY
  }
  {
    ...
  }
  {
    M_i \CONVDT N_i
  }
  {
    ...
  }
  {
    \ASTAPPLY{L, \VEC{M}} \CONVDT \AST{t}(\VEC{N})
  }{Promote dl 1}
  \\\\
  \FIVEPREMISERULENAMEDRIGHT
  {
    L \CONVDT \TAGAPPLY
  }
  {
    M \CONVDT \TAG{t}
  }
  {
    ...
  }
  {
    N_i \CONVDT N_i'
  }
  {
    ...
  }
  {
    \ASTAPPLY{L, M, \VEC{N}} \CONVDT \ASTAPPLY{\TAG{t}, \VEC{N'} }
  }{Promote dl 2}
  \\\\  
\THREEPREMISERULENAMEDRIGHT
  {
    ...
  }
  {
    M_i \CONV A_i
  }
  {
    ...
  }
  {
    \ASTAPPLY{\TAG{t}, \VEC{M}} \CONV \ASTAPPLY{\TAG{t}, \VEC{A}}
  }{Promote}
\end{RULES}

\caption{Additional rules defining $\CONVCT$, $\CONVDT$ and $\CONV$
  for AST promotion.}\label{figure_reductions_promote}

\end{FIGURE}

%% file: figure_reductions_mu.tex
\begin{FIGURE}
\begin{RULES}
	\ONEPREMISERULENAMEDRIGHT
	{
		M \CONVMU A
	}
	{
		\UPML{M} \CONVCT A
	}{UpML ct}
    \quad
	\ONEPREMISERULENAMEDRIGHT
	{
		M \CONVCT A
	}
	{
		\DOWNML{M} \CONVMU A
	}{DownML ul}
        \\\\
	\ZEROPREMISERULENAMEDRIGHT
	{
		"x" \CONVMU \ASTSTRING{"x"}
	}{String ul}
        \quad
	\TWOPREMISERULENAMEDRIGHT
	{
		M \CONVMU A		
	}
	{
		N \CONVMU B
	}
	{
		M N  \CONVMU \ASTAPP{A}{B}
	}{App ul}
        \\\\
	\ONEPREMISERULENAMEDRIGHT
	{
		M \CONVMU A		
	}
	{
		\mu g.\lambda x.M  \CONVMU \ASTREC{\ASTSTRING{"g"}}{\ASTSTRING{"x"}}{A}
	}{Rec ul}
        \\\\
	\ONEPREMISERULENAMEDRIGHT
	{
		M \CONVMU A		
	}
	{
		\lambda x.M  \CONVMU \ASTLAM{\ASTSTRING{"x"}}{A}
	}{Lam ul}
        \quad
        \ZEROPREMISERULENAMEDRIGHT
        {
                \TAG{t} \CONVMU \TAG{t}
        }{Tag ul}
        \\\\
	\ONEPREMISERULENAMEDRIGHT
	{
		M \CONVMU A		
	}
	{
		\EVAL M  \CONVMU \ASTEVAL A
	}{Eval ul}
        \quad
	\TWOPREMISERULENAMEDRIGHT
	{
		M \CONVMU A
	}
        { A \CONVMU B }
	{
		\UPML{M} \CONVMU B
	}{UpML ul}
        \\\\
	\ZEROPREMISERULENAMEDRIGHT
	{
		x \CONVMU \ASTVAR{"x"}
	}{Var ul}
\quad
        \THREEPREMISERULENAMEDRIGHT
        {
          ...
        }
        {
          M_i \CONVMU A_i
        }
        {
          ...
        }
        {
          \AST{t}(\VEC{M}) \CONVMU \ASTAPPLY{\TAG{t}, \VEC{A}}
        }{Ast ul}
\end{RULES}
\caption{Additional rules for upMLs.}\label{figure_reduction:mu}
\end{FIGURE}

%% file: figure_substitution_letdownmls.tex
\begin{figure*}[t]
\begin{RULES}
   \THREEPREMISERULENAMEDRIGHT{
     M \CONVCT A}{
     A \CONV B}{
     N\SUBST{B}{x} \CONVCT C}{
     \LETCT{x}{M}{N} \CONVCT C}{Let ct}
\\\\\\
\begin{array}{rcl}
  \DOWNML{M}\SUBST{N}{x} &=& \DOWNML{M\SUBST{N}{x}} 
     \\[1mm]
  \UPML{M}\SUBST{N}{x} &=& \UPML{M\SUBST{N}{x}} 
     \\[1mm]
  (\LETCT{x}{M}{N})\SUBST{L}{y}
    &= &
  \begin{cases}
    \LETCT{x}{M\SUBST{L}{y}}{N\SUBST{L}{y}} & x \neq y \\
    \LETCT{x}{M}{N} & x = y
  \end{cases}
\end{array}
\end{RULES}
\caption{The additional reduction rule, as well as the substitution rules, for letdownMLs. }
\label{substitution_letdownmls}
\end{figure*}

%% file: figure_reductions_lift.tex
\begin{FIGURE}
\begin{RULES}
  \ONEPREMISERULE
  {
    M \CONVCT N
  }
  {
    \LIFT{M} \CONVCT \LIFT{N}
  }
  \quad
  \ONEPREMISERULE
  {
    M \CONVDT N
  }
  {
    \ASTLIFT{M} \CONVDT \LIFT{N}
  }
  \quad
  \ONEPREMISERULE
  {
    M \CONVMU N
  }
  {
    \LIFT{M} \CONVMU \ASTLIFT{N}
  }
  \\\\
  \ONEPREMISERULE
  {
      c\ \textrm{is an integer}
  }
  {
      \LIFT{c} \CONV \ASTINT{c}
  }
  \quad
  \ONEPREMISERULE
  {
      c\ \textrm{is a string}
  }
  {
      \LIFT{c} \CONV \ASTSTRING{c}
  }
\end{RULES}
\caption{Additional rules for lifting. For simplicity, we only define lifting
for integers and strings, but one can define lifting for any type
desired.}\label{figure_reductions_lift}
\end{FIGURE}

%% file: generalFramework.tex
\section{A recipe for creating HGMP calculi}\label{generalFramework}

\newcommand{\oldcalc}[0]{$L$\xspace}
\newcommand{\newcalc}[0]{$L_{\mathsf{mp}}$\xspace}

Nothing in the presentation of our calculus  has 
relied on the $\lambda$-calculus as starting point. In this
section, we build upon an observation from
Converge that HGMP can easily be detached from the `base' language it has been
added to~\cite{tratt__compile_time_meta_programming_in_a_dynamically_typed_oo_language}:
we informally show how one can add HGMP features to a
typical programming language.

Let us imagine that we have a language \oldcalc which we wish to extend
with HGMP features to create \newcalc. We assume that \oldcalc
has its syntax given as an algebraic signature (with an indication of bindings),
and the semantics as a rule system over the syntax. We require \oldcalc to have a
string-esque datatype to represent variables; such a datatype can be trivially
added to \oldcalc if not present. We can then create \newcalc as
follows:

\begin{itemize}
  
\item Mirror every syntactic element of \oldcalc with an AST and a tag.

\item Add $\CONSTEVAL$, $\textsf{ast}_{\textsf{promote}}$ and their corresponding tags.
    
\item Add upMLs, downMLs, and letdownMLs.

\item Add appropriate reduction rules for ASTs, upMLs, downMLs, and letdownMLs.

\end{itemize}

\NI Semi-formally, we define \newcalc's syntax as follows. Assuming that
$C$ is the set of \oldcalc's program constructors, $L_{\mathsf{mp}}$'s constructors and tags
are defined as follows:
\begin{align*}
   T
      &=
   C \cup \{ \CONSTEVAL, \CONSTAPPLY \} \\[1mm]
  C_{mp}
     &=
  C\
     \cup\
  \{\CONSTEVAL, \DOWNML{\_}, \UPML{\_}, \mathsf{let}_{\downarrow}\} 
     \cup 
  \{ \AST{t}\ |\ t \in T\} \cup \{\TAG{t}\ |\ t \in T\}
\end{align*}
The arities and binders of the new syntax are as follows:
\begin{itemize*}

\item If $c \in C$ then its arity and binders are unchanged in $C_{\mathsf{mp}}$.
\item $\AST{c}$ has the same arity as $c \in C$ and no binders.
\item $\AST{\CONSTAPPLY}$ has variable arity, or, equivalently has
  arity 2, with the second argument being of type list. There are no
  binders.
\item $\AST{\CONSTEVAL}$ has arity 1 and no binders.
\item $\TAG{t}$ has arity 0 and no binders for $t \in T$.
\item $\CONSTEVAL$, $\DOWNML{\_}$, and $\UPML{\_}$ have arity 1 and no binders.
\item $\mathsf{let}_{\downarrow}$ has arity 3 and its first argument is binding.

\end{itemize*}

\noindent \newcalc inherits all of \oldcalc's reduction rules.  \newcalc's
$\CONV$ reduction relation is then augmented with the following rules:
\begin{RULES}
  \ONEPREMISERULE
  {
    t \in T
  }
  {
    t \CONV t
  }
  \qquad
  \THREEPREMISERULE
  {
    L \CONV M
  }
  {
    M \CONVDT N
  }
  {
    N \CONV N'
  }
  {
    \EVAL{L} \CONV N'
  }
  \qquad
  \FOURPREMISERULE
  {
    ...
  }
  {
    M_i \CONV N_i
  }
  {
    ...
  }
  {
    t \in T
  }
  {
    \AST{\mathrm{t}}(\VEC{M}) 
       \CONV
    \AST{\mathrm{t}}(\VEC{N}) 
  }
\end{RULES}
\NI A $\CONVDT$ relation must then be added to \newcalc. The definition of this relation 
follows the same pattern as that of $\CONVDT$ in the calculus
presented earlier in the paper: each \oldcalc constructor $c$ must have a rule
in $\CONVDT$ to convert it from an AST to a normal calculus term. If a
constructor $c$ has no binders, the corresponding rule is simple:
\[
   \THREEPREMISERULE
   {
     ...
   }
   {
     M_i \CONVDT N_i
   }
   {
     ...
   }
   {
     \AST{c}(\VEC{M}) \CONVDT c(\VEC{N})
   }
\]
Two examples of such rules are as follows:
\[
  \ZEROPREMISERULE
  {
    \ASTVAR{"x"} \CONVDT x
  }
  \qquad
  \ZEROPREMISERULE
  {
    \ASTSTRING{"x"} \CONVDT "x"
  }
\]
\noindent Constructors with binders are most easily explained by example. If
$c$ has arity 2, with the first argument being a binder, the following
rule must be added:
\[
   \TWOPREMISERULE
   {
     M \CONVDT "x"
   }
   {
     N \CONVDT N'
   }
   {
     \AST{c}(M, N) \CONVDT c(x, N')
   }
\]
\noindent The following rules must be added for higher-order ASTs:
\begin{RULES}
  \ONEPREMISERULE
  {
    M \CONVDT N
  }
  {
    \ASTEVAL{M} \CONVDT \EVAL {N}
  }
  \quad
  \THREEPREMISERULE
  {
    L \CONVDT \TAG{t}
  }
  {
    M_i \CONVDT N_i
  }
  {
    t \in T
  }
  {
    \ASTAPPLY{L, \VEC{M}} \CONVDT \AST{c}( \VEC{N} )
  }
  \\\\
  \THREEPREMISERULE
  {
    L \CONVDT \TAGAPPLY
  }
  {
    M \CONVDT \TAG{t}
  }
  {
    N_i \CONVDT R_i
  }
  {
    \ASTAPPLY{L, M, \VEC{N}} \CONVDT \ASTAPPLY{\TAG{t}, \VEC{R} }
  }
  \quad
  \ONEPREMISERULE
   {
     t \in T
   }
  {
    \TAG{t} \CONVDT \TAG{t}
  }
\end{RULES}
\NI Assuming we wish to enable compile-time HGMP, a $\CONVCT$
relation must be added:
\begin{RULES}
  \ONEPREMISERULE
  {
    M \in \{x, "x" \} \cup \{ \TAG{t} \ |\ t \in T\}
  }
  {
    M \CONVCT M
  }
  \quad
  \ONEPREMISERULE
  {
    M \CONVCT N
  }
  {
    \EVAL{M} \CONVCT \EVAL{N}
  }
  \\\\  \TWOPREMISERULE
  {
    M_i \CONVCT N_i
  }
  {
    c \in C
  }
  {
    c( \VEC{M} ) \CONVCT c( \VEC{N} )
  }
  \quad
  \TWOPREMISERULE
  {
    M_i \CONVCT N_i
  }
  {
    t \in T
  }
  {
    \AST{\mathrm{t}}( \VEC{M} ) 
       \CONVCT
    \AST{\mathrm{t}}( \VEC{N} ) 
  }  
  \\\\
  \ONEPREMISERULE
   {
     t \in T
   }
  {
    \TAG{t} \CONVCT \TAG{t}
  }
  \quad
  \THREEPREMISERULE
  {
    M \CONVCT A
  }
  {
    A \CONV B
  }
  {
    B \CONVDT C
  }
  {
    \DOWNML{M} \CONVCT C
  }
  \\\\
   \THREEPREMISERULE{
     M \CONVCT A}{
     A \CONV B}{
     N\SUBST{B}{x} \CONVCT C}{
     \LETCT{x}{M}{N} \CONVCT C}
\end{RULES}
\NI Note that the last two rules (for downMLs and letdownMLs) are added unchanged from
earlier in the paper (we assume for letdownMLs that \oldcalc has
a suitable notion of capture-avoiding substitution,
which is extended to \newcalc as described in Section
\ref{calculus}).
The rules for upMLs are given by a new relation $\CONVMU$ which is a
trivial variation of that in Figure \ref{figure_reduction:mu} and
omitted for brevity.

While semi-mechanically creating \newcalc from \oldcalc easily results in a new
language with HGMP, we cannot guarantee that \newcalc will always respect the
`spirit' of \oldcalc. For example, adding HGMP to the $\pi$-calculus in this
fashion would lead to HGMP that executes sequentially (e.g.~in the evaluation of
downMLs) which may not be desirable (although the resulting HGMPified
$\pi$-calculus would be a good starting point for developing message-passing
based forms of HGMP). Nevertheless, for most sequential programming languages,
we expect \newcalc to be in the spirit of \oldcalc. As this shows, the HGMP
features of \newcalc are easily considered separately. We suggest this helps
explain how such systems have been retro-fitted on languages such as Haskell,
and gives pointers for designers of other languages who wish to consider adding
HGMP.

%% file: typing.tex
\section{Example: staged typing and HGMP}\label{typing}

\input{figure_typing_cube}

\NI In conventional programming languages, static typing provides
compile-time guarantees that certain classes of error cannot happen at
run-time. However, HGMP blurs the lines between compile-time and
run-time, causing complications in typing that have not yet been fully
resolved. The purpose of this section is to demonstrate that our calculus can
also be useful for studying
static typing in an HGMP language. We do this by defining a type
system which is conceptually close to Template Haskell's.

\subsection{Design issues}

The three major design questions for static typing in the face of HGMP are:
what does type-safety mean in multi-staged languages? When should static types
be enforced? And: what static types should ASTs have?

Type-safety normally means that programs cannot get `stuck' at
run-time, in the sense that the program gets to a point where no reduction rules can
be applied to it, but it has not yet reached a value. The static typing system
identifies such programs and prevents them from being run.
Alas,
concepts like ``stuck'' and even ``value'' are not straightforward in
HGMP languages.  This section will only outline some of the key
issues.  We leave a detailed investigation as further work.\footnote{For example,  the computation described
  by $\CONVMU$ relation does not have a syntactic notion of value.
Consider the term $\ASTINT{3}$. Whether one
should consider it as a value with respect to $\CONVMU$ depends on whether
$\CONVMU$ was previously applied to $3$ or not.
This complicates defining a small-step semantics corresponding to
$\CONVMU$.}

There are two main choices for when static types are to be checked in
an HGMP language: upfront or in stages. Upfront typing as found in
MetaML guarantees that any program which statically
type-checks cannot get stuck in any later
stage~\cite{TahaW:mulstapitaa,TahaW:envClas}. This strong guarantee
comes at a price: many seemingly reasonable meta-programs fail to
type-check, at least for simple typing systems (e.g.~admitting type inference).  We therefore believe that -- except, perhaps, for
verification-focused languages -- staged type-checking is the more
practical of the two approaches. It guarantees only that, whenever an
AST is $\CONVDT$ converted to a normal $\lambda$-term as a result of a downML or
$\CONSTEVAL$, the program will not get stuck before the next such conversion.
Thus, type-checking might need to be carried out more than once, and the guarantees at
each stage are weaker than in upfront checking. In the rest of this paper, we only
consider staged type-checking.

There are three main ways that code (be that ASTs or MetaML-esque
quasi-quotes) can be statically typed. In a \emph{monotyped} system,
every program representing code has the same type $\CODE$. In a
\emph{parameterised} system, code of type $\TYPEDCODE{\alpha}$ can be
shifted a meta-level (at compile-time or run-time) to a program of type
$\alpha$. Finally, it is possible to bridge these two extremes with a
\emph{hybrid} system which allows both parameterised and
monotyped code types. MetaML uses
parameterised code types. Template Haskell is currently monotyped (though there are
proposals for it to move to a hybrid
system~\cite{PeytonJonesS:newdirfth}) and we thus use that as the
basis of our typing system.
Figure \ref{figure_typing_cube} surveys how different HGMP languages approach typing.

\subsection{Staged typing for the foundational calculus}\label{simpleTyping}\label{stuck programs}

The key properties in the staged typing we define are as follows:

\begin{enumerate}

\item\label{typing:rule:1} All code that is evaluated by the
    $\CONV$ relation will have been previously type-checked and thus cannot get stuck.
    This is done by type-checking the expressions inside downMLs and
    $\CONSTEVAL$s, and type-checking the complete program after all downMLs have been
    removed.

\item\label{typing:rule:2} All code that is converted by the $\CONVDT$ relation
    will have been previously type-checked to ensure that the ASTs
    involved are properly formed and thus applying $\CONVDT$ cannot get stuck.

\item\label{typing:rule:3} Since the only possible places where
    $\CONVCT$ could get stuck are where it references
    $\CONV$ and $\CONVDT$, (\ref{typing:rule:1},
  \ref{typing:rule:2}) guarantee that $\CONVCT$
  doesn't get stuck.

\item\label{typing:rule:4} Since $\CONVMU$ could only get stuck where
    it references $\CONVCT$, (\ref{typing:rule:3}) guarantees
    that $\CONVMU$ doesn't get stuck.
\end{enumerate}

\NI To make this form of typing  concrete, we create a type system for this paper's
calculus (modifying $\CONSTEVAL$ for reasons that will soon become clear). The
type system can be seen as an extension of Template Haskell's, augmented with
higher-order HGMP and run-time HGMP.    

We first need to extend the calculus grammar to introduce types as follows:
\begin{GRAMMAR}
	M
		&::=&
        ...
		\VERTICAL
	\mu g.\lambda x.M
                \VERTICAL 
        \TYPEDEVAL{\alpha}{M} 
                \\[1mm]
        t
                &::=&
        ...
		\VERTICAL
        \CONSTREC
		\VERTICAL
        \CONSTTYPEDEVAL{\alpha}
                \\[1mm]
   \alpha &::=& \INT \VERTICAL \BOOL \VERTICAL \alpha \FS \beta
   \VERTICAL \STRING \VERTICAL \CODE \VERTICAL \TAGTYPE{t}
\end{GRAMMAR}

\begin{figure}[t]
\begin{RULES}
  \ONEPREMISERULE
  {
    M \CONVCT N
  }
  {
    \TYPEDEVAL{\alpha}{M} \CONVCT \TYPEDEVAL{\alpha}{N}
  }
        \quad
  \ONEPREMISERULE
  {
    M \CONVDT N
  }
  {
    \ASTTYPEDEVAL{\alpha}{M} \CONVDT \TYPEDEVAL{\alpha}{N}
  }
  \\\\
  \ONEPREMISERULE
  {
    M \CONVMU A		
  }
  {
    \TYPEDEVAL{\alpha}{M}  \CONVMU \ASTTYPEDEVAL{\alpha}{A}
  }
\end{RULES}
\caption{$\CONVCT$, $\CONVDT$, $\CONVMU$ reduction relations for the altered $\CONSTEVAL$.}
\label{alteredevalrules}
\end{figure}

\NI We assume readers are acquainted with static types for the basic
$\lambda$-calculus. ASTs have type $\CODE$; each $\TAG{t}$ has a
corresponding static type $\TAGTYPE{t}$.  The only surprising change is the type
annotation of $\TYPEDEVAL{\alpha}{M}$ and the corresponding tag
$\CONSTTYPEDEVAL{\alpha}$.  The type annotation $\alpha$ is used for
type-checking the program $N$, obtained from $M$ by evaluation to an
AST and subsequent $\CONVDT$ conversion back to a normal $\lambda$-term: all we have to do is verify that $N$ has
type $\alpha$. Without this type annotation, we would have to type-check $N$
and ensure that $N$'s type is compatible with its context. The additional reduction rules for all relations
except $\CONV$ are shown in Figure~\ref{alteredevalrules}.

The core of the approach is to intersperse type-checking with reduction, making
sure that we can never run code that has not been type-checked.  We therefore
first add a `normal' type-checking phase between compile-time and run-time
(i.e~for programs which do not use downMLs or $\CONSTEVAL$, this phase is the
only type-check invoked):
\[
  \underbrace{M \CONVCT A}_{\text{compile-time}} \ 
  \overbrace{\TYPES{}{A}{\alpha}}^{\text{type-checking}} \ 
  \underbrace{A \CONV V}_{\text{run-time}}
\]

\NI Second, we must add a type-checking phase to ensure that code generated at
compile-time and inserted into the program by downMLs is type-safe (i.e.~the
expression in a downML has a static type of $\CODE$). We therefore alter
[\RULENAME{DownML ct}] as follows:
\[
  \FOURPREMISERULENAMEDRIGHT
  {
    M \CONVCT A
  }
  {
    \TYPES{}{A}{\CODE}
  }
  {
    A \CONV B
  }
  {
    B \CONVDT C
  }
  {
    \DOWNML{M} \CONVCT C
  }{DownML ct}
\]
\noindent Finally, we alter [\RULENAME{Eval rt}] to perform a type-check
on the code it will evaluate at run-time:
\[
  \FOURPREMISERULENAMEDRIGHT
  {
    L \CONV M
  }
  {
    M \CONVDT N
  }
  {
    \TYPES{}{N}{\alpha}
  }
  {
    N \CONV N'
  }
  {
    \TYPEDEVAL{\alpha}{L} \CONV N'
  }{Eval rt}
\]

\NI We define the typing judgement
$\TYPES{\Gamma}{M}{\alpha}$ as follows. $M$ is a program that does not
contain upMLs or downMLs (which will have been removed by $\CONVCT$
before type-checking). $\Gamma$ is an environment (i.e.~a finite map)
from variables to types such that all of $M$'s free
variables are in the domain of $\Gamma$. Note that type-checking
needs to be applied only to programs without downMLs and upMLs, hence the free variables of
a program not containing upMLs and downMLs can be defined as usual
for a program, and we omit the details.
We write $\Gamma, x : \alpha$ for the typing
environment that extends $\Gamma$ with a single entry, mapping $x$ to
$\alpha$, assuming that $x$ is not in $\Gamma$'s domain.  We write
$\TYPES{}{M}{\alpha}$ to indicate that the environment is empty.

\input{figure_typing_nonParameterised}

The rules defining $\TYPES{\Gamma}{M}{\alpha}$ are given in Figure
\ref{figure_typing_nonParameterised}. The rules for variables,
function abstraction, recursion, and application are as in
conventional $\lambda$-calculus. For ASTs $\AST{t}(\VEC{M})$ where $t$
is not one of $\CONSTLAM$, $\CONSTREC$, $\CONSTAPPLY$, $\CONSTSTRING$,
or $\CONSTINT$, if all the arguments have type $\CODE$ then
$\AST{t}(\VEC{M})$ also has type $\CODE$. ASTs representing
a binding construct (e.g.~$\ASTLAM{M}{N}$) have type $\CODE$ if: the terms
representing binders are of the form $\ASTSTRING{L}$ with $L$ having type
$\STRING$; and $N$ has type $\CODE$.

\subsection{Examples}

With the typing system defined, a few examples can help understand how and when
it operates. First we note that terms such as $\ASTLAM{(\lambda x.x)}{M}$ that
would get stuck without the type system do not type-check in our system. Second
we can see that some expressions pass one type-check and fail a later one. Consider
the following program:
  \[
      2\ + \DOWNML{\ \ASTLAM{\ASTSTRING{"x"}}{\ASTVAR{"x"}}\ }
  \]
The downML $\CONVCT$ reduces to $\ASTLAM{\ASTSTRING{"x"}}{\ASTVAR{"x"}}$ which
successfully type-checks as being of type $\mathsf{code}$. The entire program
then $\CONVCT$-reduces to $2\ + \lambda x.x$, which is neither a value nor has
any $\CONV$-reductions and fails to type-check.

The type system can also check more complex properties, such as AST constructors
with the wrong number of arguments. Let $M$ be the program $ \ASTAPPLY{ \TAGINT,
\ASTINT{1} }$ in the following program:
  \[
     \ASTAPPLY{
        \TAGAPPLY,
        \TAGINT,
        M,
        M}
  \]
When run through a downML or $\CONSTEVAL$ for the first time it will yield:
  \[
     \ASTAPPLY{\TAGINT,  \ASTINT{1},  \ASTINT{1} }
  \]
which type-checks correctly. However, if this AST is run through a downML or
$\CONSTEVAL$ it results in:
  \[
     \ASTINT{1, 1}
  \]
which fails to type-check.

%% file: figure_typing_cube.tex
\begin{figure*}[t!]
\centering
\begin{tabular}{lccccccc}
\toprule
Language         & Monotyped & Parameterised & Hybrid & Dynamic & Staged & Upfront\\
\midrule
Converge         & $\circ$   & $\circ$   & $\circ$ & $\bullet$ & n/a & n/a \\
Lisp             & $\circ$   & $\circ$   & $\circ$ & $\bullet$ & n/a & n/a \\
MetaML           & $\circ$   & $\bullet$    & $\circ$        & $\circ$          & $\circ$ & $\bullet$ \\
Template Haskell & $\bullet$ & $\circ$   & $\circ$    & $\circ$       & $\bullet$ & $\circ$ \\
Scala (\texttt{scala.meta}) & $\bullet$ & $\circ$   & $\circ$    & $\circ$       & $\bullet$ & $\circ$ \\
\bottomrule
\end{tabular}
\caption{How different HGMP languages approach typing.}
\label{figure_typing_cube}
\end{figure*}

%% file: figure_typing_nonParameterised.tex
\begin{FIGURE}
\begin{RULES}
  \ONEPREMISERULE
  {
    \TYPES{\Gamma, x : \alpha}{M}{\beta}
  }
  {
    \TYPES{\Gamma}{\lambda xM}{\alpha \FS \beta}    
  }
  \quad
  \ONEPREMISERULE
  {
    \TYPES{\Gamma, g : \alpha \FS \beta, x : \alpha}{M}{\beta}
  }
  {
    \TYPES{\Gamma}{\mu g.\lambda x.M}{\alpha \FS \beta}    
  }
  \quad
  \ZEROPREMISERULE
  {
    \TYPES{\Gamma, x : \alpha}{x}{\alpha}
  }
  \\\\
  \TWOPREMISERULE
   {
     \TYPES{\Gamma}{M}{\alpha \FS \beta}
   }
   {
     \TYPES{\Gamma}{N}{\alpha}
   }
   {
     \TYPES{\Gamma}{MN}{\beta}
   }
  \quad
  \TWOPREMISERULE
   {
     \TYPES{\Gamma}{M}{\STRING}
   }
   {
     \TYPES{\Gamma}{N}{\CODE}
   }
   {
     \TYPES{\Gamma}{\ASTLAM{\ASTSTRING{M}}{N}}{\CODE}
   }
   \\\\
     \ZEROPREMISERULE
   {
     \TYPES{\Gamma}{\TAG{t}}{\TAGTYPE{k}}
   }
   \quad
   \THREEPREMISERULE
   {
     \TYPES{\Gamma}{L}{\STRING}
   }
   {
     \TYPES{\Gamma}{M}{\STRING}
   }
   {
     \TYPES{\Gamma}{N}{\CODE}
   }
   {
     \TYPES{\Gamma}{\ASTREC{\ASTSTRING{L}}{\ASTSTRING{M}}{N}}{\CODE}
   }
   \\\\
   %% \quad
   %% \ONEPREMISERULE
   %% {
   %%   \TYPES{\Gamma}{M}{\CODE}
   %% }
   %% {
   %%   \TYPES{\Gamma}{\TYPEDEVAL{\alpha}{M}}{\alpha}
   %% }
   %% \quad
   %% \ONEPREMISERULE
   %% {
   %%   \TYPES{\Gamma}{M}{\alpha}
   %% }
   %% {
   %%   \TYPES{\Gamma}{\UPML{M}}{\CODE}
   %% }
   %% \quad
   %% \ONEPREMISERULE
   %% {
   %%   \TYPES{\Gamma}{M}{\CODE}
   %% }
   %% {
   %%   \TYPES{\Gamma}{\INSERT{M}}{\CODE}
   %% }
   \quad
   \FOURPREMISERULE
   {
     t \neq \CONSTLAM, \CONSTREC, \CONSTAPPLY, \CONSTSTRING, \CONSTINT
   }
   {
     ...
   }
   {
     \TYPES{\Gamma}{M_i}{\CODE}
   }
   {
     ...
   }
   {
     \TYPES{\Gamma}{\AST{t}{(\VEC{M})}}{\CODE}
   }
   \\\\
   \ONEPREMISERULE
   {
     \TYPES{\Gamma}{M}{\INT}
   }
   {
     \TYPES{\Gamma}{\ASTINT{M}}{\CODE}
   }
   \quad
   \ONEPREMISERULE
   {
     \TYPES{\Gamma}{M}{\STRING}
   }
   {
     \TYPES{\Gamma}{\ASTSTRING{M}}{\CODE}
   }
   \\\\
   \ONEPREMISERULE
   {
     \TYPES{\Gamma}{M}{\CODE}
   }
   {
     \TYPES{\Gamma}{\TYPEDEVAL{\alpha}{M}}{\alpha}
   }
   \quad
   \FOURPREMISERULE
   {
     \TYPES{\Gamma}{M}{\TAGTYPE{t}}
   }
   {
     ...
   }
   {
     \TYPES{\Gamma}{N_i}{\CODE}
   }
   {
     ...
   }
   {
     \TYPES{\Gamma}{\ASTAPPLY{M, \VEC{N}}}{\CODE}
   }
\end{RULES}
\caption{Type-checking with type $\CODE$ for programs not containing
  upMLs and downMLs. Some straightforward cases
  omitted.}\label{figure_typing_nonParameterised}
\end{FIGURE}

%% file: conclusion.tex
\section{Related work}

Meta-programming is such a long-studied subject that a full related work section
would be a paper in its own right. We have referenced many real-world systems in
previous sections; in this section, we therefore concentrate on related work
that has a foundational or formal bent, and which has not been previously mentioned.

Run-time HGMP has received more attention than compile-time HGMP, with
MetaML and the \textit{reFL$^{\!\mbox{\text{ect}}}$}
language~\cite{GrundyJ:reffunlfhdatp} being amongst the well known
examples. MetaML is the closest in spirit to this paper, though it has
two major, and two minor, differences. The minor differences are that MetaML is
typed and hygienic, whereas our system can model untyped and non-hygienic
systems, enabling people to experiment with different notions of each. The first
major difference is that MetaML does not model compile-time evaluation
of arbitrary code (see below for a discussion of MacroML, which partly addresses
this). The second major difference is upMLs and ASTs: MetaML has only the
former, while our system has both, with ASTs the `fundamental' construct and
upMLs a convenience atop them. As discussed in Section~\ref{upmls}, this
restricts the programs -- and hence programming languages -- that can be
expressed.

Run-time HGMP is also the primary object of study in \emph{unstaging translations}
(see e.g.~\cite{ChoiW:staanaomspvut,InoueJ:stabeytpac,KameyamaY:clostafscttc}).
These are semantics-preserving embeddings of an HGMP language
into a language without explicit
constructs for representing code as data. Depending on the particular
pairing of source and target language, the unstaging translation
can be extremely complex, making it difficult to use as a mechanism
for understanding the fundamental constructs. We are also not aware of unstaging translations that
treat compile-time and run-time HGMP in a unified way.  An
open research question is whether our general approach in
Section~\ref{generalFramework} can be unstaged in a generic way.

Compile-time HGMP research has mostly focused on Lisp macros
(e.g.~\cite{bove92confluent,HermanD:thehygmac}) or C++ templates
(e.g.~\cite{GarciaR:towfouftrm}). Perhaps the work most similar to ours
is the formal model of a large subset of Racket's macro
system~\cite{flatt12macros}. However, this formalises Racket's
\texttt{define-syntax} system which is dynamically typed, and not a HGMP system in our definition
(see Section~\ref{hgmp vs macro expansion}). The system we define
is closer in spirit to a statically typed version of Racket's \texttt{syntax-case} system.
MacroML~\cite{GanzS:macmulsc} investigates Lisp-style macro systems by translation into  MetaML.
The key insight is that macros are special constructs which must be
entirely expanded in a separate stage before any normal code is evaluated. Our approach
instead models systems where normal code can be evaluated at both
compile-time and run-time.

Research on types for HGMP and the relationship with modal
logics via a Curry-Howard correspondence started with work by Davis
and Pfenning~\cite{DaviesR:modanaosc,DaviesR:temlogatbta}. In recent
years, more expressive typing systems along these lines have been investigated
(see
e.g.~\cite{NanevskiA:conmodtt,TsukadaT:logfoufecLONG}).  The axiomatic
semantics of HGMP is explored in~\cite{BergerM:prologfhml}. Some
original approaches towards the foundations of run-time HGMP are:
M-LISP~\cite{muller92mlisp} which provides an operational semantics
for a simplified Lisp variant with $\mathsf{eval}$ but without macros;
Archon~\cite{StumpA:dirrefmp}, which is based on the untyped
$\lambda$-calculus but without an explicit representation of
code; the two-level $\lambda$-calculus \cite{GabbayMJ:twolevlc} which
is based on nominal techniques; and the $\rho$-calculus
\cite{MeredithLG:refhigoc} which combines ideas from Conway games and
$\pi$-calculus. 

Issues closely related to HGMP have been studied in the field of logic,
often under the heading of reflection~\cite{HarrisonJ:metrefitpasac}.  Little work seems to have been done
towards unification of the multiple approaches to meta-programming.
Farmer et al.'s concept of syntax frameworks~\cite{FarmerWM:forsynbmauqae,FarmerWM:fraforrastuqae} may well have
been the first foray in this direction but are not
intended to be a full model of meta-programming, whether homogeneous or heterogeneous. In particular, they
do not capture the distinction between compile-time HGMP and run-time
HGMP.

\section{Conclusions}
\label{conclusions}

In this paper we presented the first foundational calculus for modelling
compile-time and run-time HGMP as found in languages such as Template Haskell.
The calculus is designed to be considered in increments, and adjusted as needed
to model real-world languages. We provided a type system for the calculus.
We hope that the calculus provides a solid
basis for further research  into HGMP. 

The most obvious simplification in the calculus is its treatment of names. The
calculus deliberately allows capture and is not hygienic since there are different
styles of hygiene, and various possible ways of formalising it.
A system similar to Template Haskell's, for example, where names in upMLs are
preemptively $\alpha$-renamed to fresh names would be a simple addition, but other,
sometimes more complex, notions are possible (e.g.~determining which variables
should be fresh and which should allow capture). We hypothesise that
the formal definition of hygiene in~\cite{adams15hygiene}, which is based on  
nominal techniques \cite{PittsAM:nomsetnasics}, can be adapted to our foundational
calculus.

%% --------------

%% Recently several papers have investigated  two aspects of HGMP:
%% \begin{itemize}

%% \item Hygiene: \cite{adams15hygiene,FlattM:binasetos}.

%% \item Types: Serveral key themes.

%% \begin{itemize}

%% \item Open code, e.g.~\cite{KimIS:polmodtsfllmsl}.
%% \item Polymorphism.   
%% \item Control effects.
%% \item Curry-Howard correspondence.
%% \item Typing the embedding of DSLs (e.g.~lightweight modular staging \cite{RompfT:ligmods},
%% a library-based approach to multistage programming uses types, rather
%% than quasi-quotes or ASTs, to distinguish between binding times).

%% \end{itemize}
%% All the above work focus on RTMP, and on quasi-quotes as sole form of
%% representing programs as data.  We believe that the techniques
%% complement our work, and can be made to work for CTMP too, and also be
%% generalised to ASTs, using our translation.

%% \item Convenient APIs (e.g. \texttt{scala.meta}).

%%  This paper extends the
%% multi-stage typing systems
%% of \cite{DaviesR:temlogatbta,DaviesR:modanaosc} so that code types
%% contain typing environments. For example open code such as
%% $\UPML{x+1}$ could be given type $\TYPEDCODE{x : \INT \vdash \INT}$,
%% where the typing environment $x : \INT$ left of the turnstile
%% describes the free variables in the code. Use of code types with
%% environments is made convenient by adding row variables a la R\'emy,
%% and ML-style let-polymorphism. Surprisingly, despite it's considerable
%% expressive power, the typing system enjoys full inference of principal
%% types.

%% \end{itemize}

\subsubsection*{Acknowledgements.} We thank W.~Farmer, A.~Kavvos, O.~Kiselyov,
G.~Meredith, S.~Peyton Jones, M.~Stay and L.~T.~van Binsbergen for
discussions about meta-programming, and E.~Burmako for answering
questions about \texttt{scala.meta}.
Laurence Tratt was funded by the EPSRC `Lecture' fellowship (EP/L02344X/1).